# Exciton Transport in Disordered Perovskite Nanocrystal Solids

Simon F. Solari[1,2], Enrique Arévalo Rodríguez[1,2], Antonella Cutrupi[1,2], Amalia Coro[1,2,3], Marc Meléndez[1,2], Alicia De Andrés[3], Almudena Torres-Pardo[4], Beatriz H. Juárez[3], and Ferry Prins[1,2*]

[1] Condensed Matter Physics Center, IFIMAC, Universidad Autónoma de Madrid, Madrid 28049, Spain

[2] Departamento de Física de la Materia Condensada, Facultad de Ciencias, Universidad Autónoma de Madrid, Madrid 28049, Spain

[3] Instituto de Ciencia de Materiales de Madrid, ICMM, Consejo Superior de Investigaciones Científicas, CSIC, Madrid 28049, Spain

[4] Departamento de Química Inorgánica, Facultad de Ciencias Químicas, Universidad Complutense de Madrid, Madrid 28040, Spain

*All correspondence should be addressed to ferry.prins@uam.es

## Abstract

**Solution-processed thin films of colloidal lead halide perovskite (LHP) nanocrystals (NCs) show great potential for the implementation into optoelectronic devices such as light-emitting diodes (LEDs), lasers, and solar cells. However, these hybrid LHP NC solids exhibit non-negligible size and shape polydispersity, which introduces both structural and energetic disorder. Here, we resolve the exciton dynamics in space, time, and energy to elucidate the impact of different forms of disorder (structural and energetic) on exciton transport. We show that the disorder depends sensitively on the length of the alkylamine ligand used in the synthesis. While shorter alkyl chain lengths lead to high polydispersity, longer alkyl chains lead to more monodispersed and smaller particles where quantum confinement becomes more pronounced and, consequently, lead to increased energetic disorder. Strikingly, we find that exciton transport is less efficient in NC solids with long alkyl chain ligands, despite having a significantly more monodisperse ensemble. This demonstrates that energetic disorder, rather than structural disorder, is the dominant factor for predicting exciton transport within these materials. These findings reveal the critical role of ligand engineering in designing high-performance optoelectronic devices based on hybrid LHP NCs, providing new insights into energy transport dynamics in disordered systems and highlighting the versatility of these materials for advanced photonic and optoelectronic applications.**

Colloidal nanocrystals (NCs) of hybrid lead halide perovskites (LHPs) with the general formula $APbX_3$, where A consists of an organic cation (methylammonium, $CH_3NH_3^+$ or formamidinium, $CH(NH_2)_2^+$) and X are halides ($Cl^-$, $Br^-$, $I^-$), have emerged as an interesting class of semiconductor materials due to their unique optical properties, such as bright emission, photoluminescence high quantum yield (PLQY), and wide color tunability.[1-9] In addition to their excellent optoelectronic properties,[10] the low-cost solution-processed films of hybrid LHP NCs show great potential to be implemented into light-harvesting photovoltaic devices, as well as photon-emitting technologies such as light-emitting diodes (LEDs) and lasing applications.[11-18]

From a synthetic point of view, and in contrast to the all-inorganic $CsPbX_3$ counterparts that are synthesized at high temperatures,[19, 20] colloidal hybrid LHP NCs can be prepared at room temperature employing the ligand-assisted reprecipitation (LARP) technique.[21, 22] Kumar *et al.* showed a room-temperature synthesis of hybrid $APbBr_3$ NC solids, where A represents a 50:50 mixture of the organic methylammonium and formamidinium cations, passivated by a series of organic alkylamine ligands.[23] By modifying the length of the saturated hydrocarbon chain of amphiphilic ligands, the shape and size distribution of green-emitting NCs, and hence their photophysical properties, could be adjusted. Specifically, longer alkyl chains lead to a decrease in particle size and an energetic blueshift of the photoluminescence (PL) emission due to a higher degree of quantum confinement.[23]

Crucially though, the lack of temporal separation of nucleation and growth stages of the NCs in the LARP approach limits the control over the shape and size distribution of the perovskite NCs. Unlike hot injection,[24] where rapid precursor mixing at elevated temperature effectively decouples nucleation from growth, LARP operates under ambient conditions where nucleation and growth stages cannot be separated in time.[21] This inherently reduces the degree of size-focusing achievable for a given ligand alkyl chain length, resulting in broader size distributions and less precise control over NC dimensions. This is particularly true for short alkyl chain-length ligands, where less effective passivation and more dynamic ligand adsorption–desorption may promote heterogeneous growth and ripening, thereby broadening the size distribution of NC solutions, which show increased disorder when processed into thin films.[2, 21, 25] Nevertheless, despite the presence of disorder, high-performance LEDs with emissive layers based on these materials have been successfully demonstrated with external quantum efficiencies of up to 25%.[26, 27]

While significant attention has been given to the study of exciton transport in highly ordered and monodisperse ensembles of perovskite NCs,[28-33] studies describing the energy transport in disordered perovskite NC with increased polydispersity are lacking.[34] Size polydispersity in NC solids leads in first place to structural disorder due to variations in size, shape, interparticle spacing, and orientation of the constituent NCs.[28] The Förster-like non-radiative energy transfer processes (NRET) that mediate energy transport are highly sensitive to interparticle spacing, causing exciton transfer rates to vary accordingly.[35, 36] Additionally, in the presence of quantum confinement, size polydispersity leads to energetic disorder.[28] The increase in the effective bandgap for smaller particles within the distribution causes site-to-site energetic disorder which favors downhill exciton migration and leads to a progressive slowing down of transport as excitons reach the lower energy sites in the ensemble.[37] Importantly, perovskite NCs are generally characterized by weak quantum confinement effects where the coexistence of structural and energetic disorder highly depends on the details of the particle distribution.[38, 39] To date, the detailed mechanisms of the interplay between energetic and structural disorder and their influence on exciton transport in disordered perovskite NC solids are not fully understood.

In this work, we present a study of exciton diffusion in room-temperature synthesized organic-inorganic hybrid LHP NC solids passivated by a series of amphiphilic alkyl amine ligands commonly used in optoelectronic devices.[17, 23, 26, 27] We report a varying extent of disorder in the system depending on the alkyl chain length of the corresponding ligand used in the colloidal synthesis. Short chain lengths lead to larger and more polydisperse NCs that exhibit significant structural disorder, while longer chain lengths lead to smaller and more monodisperse NCs whose properties are increasingly dictated by quantum confinement. Somewhat counterintuitively, time-resolved spectroscopy shows that more monodisperse ensembles exhibit larger degrees of energetic disorder, which can be attributed to an increase in quantum confinement that amplifies the bandgap variations and the site-to-site disorder for the smaller NCs. We use time-resolved microscopy to directly visualize exciton diffusion in the different NCs solids (thin films). Our results show that increased energetic disorder leads to a rapid reduction in transport efficiency by trapping excitons in low-energy sites, thus acting as the main limiting factor for exciton transport. Our results are supported by numerical simulations, revealing a complex interplay between the relative roles of energetic and structural disorder in exciton transport. They also highlight the importance of the right selection of ligands in semiconductor NC solids for the design of future high-performance optoelectronic devices with optimized energy transport properties.

Colloidal $CH_3NH_3PbBr_3$ NCs are synthesized at room temperature following the solution-based LARP protocol.[21] **Figure 1a** presents a simplified reaction scheme for the LARP process carried out in this work (for details, see Supporting Information). In short, the precursor salts $PbBr_2$ and $CH_3NH_3Br$ are dissolved in polar solvents (*N*,*N*-dimethylformamide and ethanol) and mixed with toluene containing oleic acid and a series of amphiphilic alkylamine ligands.[23] Here we study three different types of primary alkylamine ligands with different alkyl chain length, *n*, including octylamine (C8), dodecylamine (C12), and hexadecylamine (C16). To minimize variation in quality and production yield of the NCs, the reaction system is degassed with argon before and right after the addition of the precursor solutions.[5] After work-up, the colloidally dispersed NCs are purified with methyl acetate as anti-solvent to remove the unreacted precursors and the excess of ligands used for the synthesis.[4,40] The room-temperature LARP approach yields colloidal perovskite NCs of different sizes and shapes (nanocubes and nanoplatelets) with relatively large size distributions, passivated by organic ligands (see **Figure 1b**).[41] The colloidally dispersed NCs synthesized *via* the LARP approach yield strong green-emitting fluorescence under UV excitation (**Figure S1**). PLQY values for the colloidal solutions are consistently above 70%, increasing systematically with ligand length, from 77% (C8) to 84% (C12) and 97% (C16) (see **Table S1** for details).

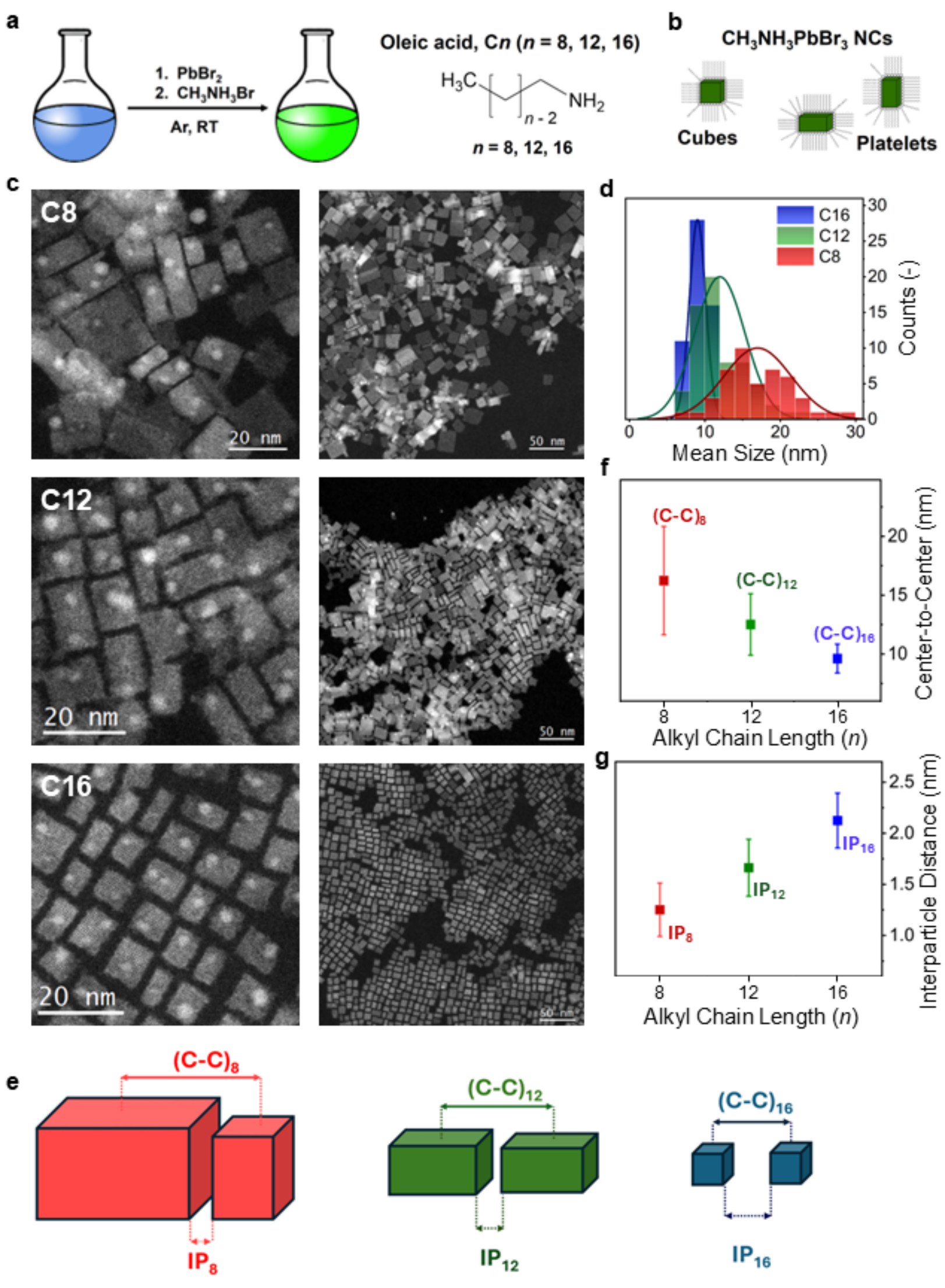

Figure 1. Synthesis and morphology of hybrid LHP NCs. (a) Simplified scheme for the synthesis of colloidal $CH_3NH_3PbBr_3$, the corresponding ligands used in the synthesis, and the molecular structure of the amphiphilic alkyl amine ligands, where *n* denotes the number of carbon atoms in the alkyl chain. (b) Illustration of colloidal $CH_3NH_3PbBr_3$ NCs of different size and shape synthesized *via* the LARP approach and passivated by organic ligands ranging from nanocubes to nanoplatelets. (c) Representative STEM micrographs at different magnifications for $CH_3NH_3PbBr_3$ NCs synthesized with C8, C12, and C16, respectively. The appearance of bright, rounded contrasts on the surfaces of the particles suggests the formation of metallic Pb as a result of exposure to the high-energy electron beam. (d) Corresponding mean size analysis showing increasing degree of polydispersity with decreasing ligand length. (e) Simplified scheme of two neighboring NCs demonstrating the role of center-to-center (C-C) and interparticle (IP) distance, $d_{C\text{-}C}$ and $d_{IP}$, respectively, as a function of the alkyl chain length in a typical NC solid. (f) Analysis of $d_{C\text{-}C}$ and (g) $d_{IP}$ as a function of the alkyl chain length.

To determine the size and shape distribution, we perform scanning transmission electron microscopy (STEM) at different magnifications of the $CH_3NH_3PbBr_3$ NCs synthesized using three different amphiphilic alkylamine ligands (C8, C12, and C16), as shown in **Figure 1c**. Upon increasing the amount of carbon atoms in the ligand chain, we observe a gradual transition from nanoplatelets to nanocubes together with a significant decrease in lateral sizes (from around 20 nm to below 10 nm).[23] Size analysis of the STEM images (details in Supporting Information and **Figure S2**) reveal the increasing polydispersity upon decreasing the length of the alkyl amine ligand ranging from 12% to approximately 27% (**Figure 1d**). Consistent with previous literature, this result evidences that the chain length of the alkylamine ligands plays an important role in controlling the size and shape of colloidal perovskite NCs synthesized *via* the LARP technique at room temperature.[42, 43] Longer ligands increase steric repulsion and hydrophobicity, which restricts the growth to smaller and highly monodisperse NCs. In contrast, shorter ligands facilitate the growth of larger nanostructures with increased polydispersity.[23] Translated to solid state, this means that, in general terms, structural disorder is significantly lower for NCs solids containing NCs synthesized with C16 compared to those synthesized with C8. The specific relationship between ligand length and particle size in LARP synthesis has important consequences for the morphology of the film, where interparticle distance ($d_{IP}$) and center-to-center distance ($d_{C\text{-}C}$) have opposite scaling (see **Figures 1e-g**). While the use of longer alkyl chains naturally leads to larger $d_{IP}$ values, the average $d_{C\text{-}C}$ is reduced due to the smaller NC sizes that are obtained. The understanding of the variation of these parameters across the C8-C16 series is essential to properly understand exciton diffusion in these NCs solids.

We compare the photophysical properties of the synthesized NCs in solution and solid state. Perovskite NC thin films are prepared by spin-coating the purified colloidal dispersions on glass slides (**Figure S3**). In solution, we observe a shift to higher energies in both the PL and absorption spectra upon increasing the alkyl chain length from C8 to C16 (**Figures 2a**, **b**). This can be explained by a higher degree of quantum confinement due to a reduced NC size upon elongating the length of the capping ligands, as confirmed by the STEM analysis (see Figure 1c).[42] **Figure 2c** summarizes the evolution of the maximum PL emission energy and the absorption band edge as a function of the alkyl chain length of the ligands used for the synthesis of the hybrid LHP NCs. We report values for the Stokes shift of a few meV, which are comparable to those previously reported in literature for $CH_3NH_3PbBr_3$ nanostructures.[44]

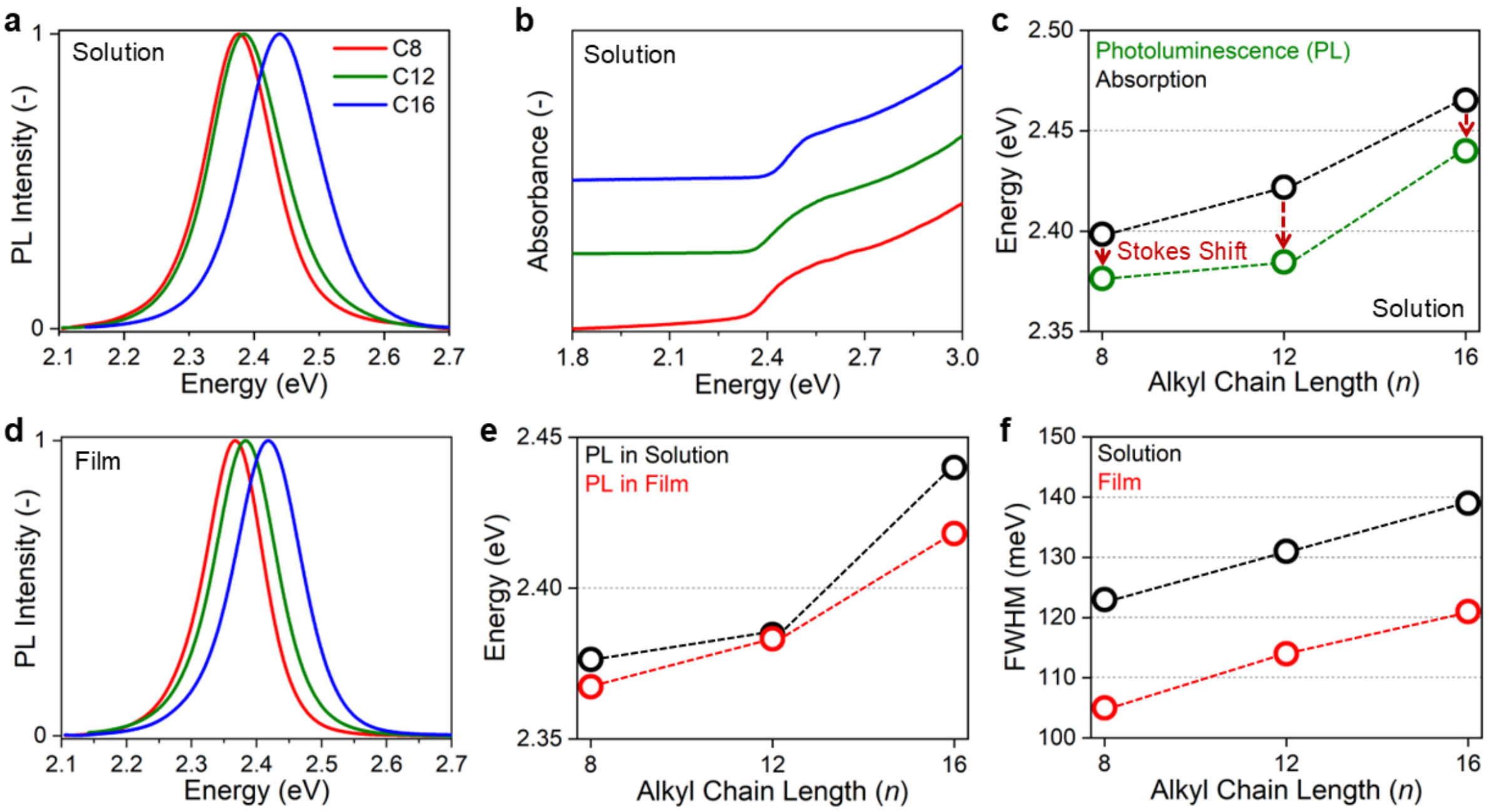

Figure 2. Photophysical properties of hybrid LHP NCs. (a) Steady-state PL and (b) absorption spectra of the colloidal solutions of $CH_3NH_3PbBr_3$ NCs with C8, C12, and C16 as capping ligands. (c) Maximum PL emission energy and the absorption band edge in solution as a function of the alkyl chain length of the ligands used for the synthesis of $CH_3NH_3PbBr_3$ NCs. (d) Steady-state PL spectra of the corresponding NC solids. (e) Comparison of PL emission energy in solution and film as a function of the alkyl chain length. (f) PL full width at half maximum, FWHM, in solution and film as a function of the alkyl chain length.

The steady-state PL spectra of the NC solids (**Figure 2d**) show similar trends as in solution where, as the alkyl chain length increases, both an energetic blueshift in the PL emission (**Figure 2e**) and an increasing PL full-width-at-half-maximum (FWHM, **Figure 2f**) are observed. The consistent observation of size-dependent emission and absorption, as well as the linear power-dependence of PL measurements for all three alkyl chain lengths (see **Figure S4**), are a clear confirmation of the excitonic nature of the optical excited state in these systems. Excitonic energy transport is mediated by dipole-dipole coupling between neighboring nanocrystals, and within the Förster approximation, the relevant distance parameter is then $d_{C-C}$,[35, 37] that is, the center-to-center distance, which as shown in Figure 1f, it is the smallest for the largest interparticle distance system (C16). Therefore, comparing purely the morphologies of the different films, the smaller and more monodisperse NC films based on C16 capping ligands would be expected to exhibit more efficient energy transport properties; however, as shown below, this is not the case.

Interestingly, as shown in Figure 2e, the thin-film emission spectra show a small energetic redshift as compared to solution. This is a first indication of exciton transport and the presence of energetic disorder, with high energy excitons transferring their energy to lower energy sites in the ensemble as they migrate through the film (a process that is absent in solution).[45, 46] To better quantify the extent of energetic disorder in the hybrid perovskite NC solids, we perform spectrally resolved transient PL measurements of exciton dynamics.[37] **Figure 3a** shows the temporal evolution of the PL spectra of the three NC solids investigated in this work. For all three samples, we observe a redshift of the median emission wavelength over the first few nanoseconds followed by its saturation at longer times. The decay dynamics of the three colloidal NCs solutions can be seen in **Figure S5**. Please note that in solution, such transient redshifts are indeed absent (see **Figure S6**). In solid state, these redshift observations are consistent with exciton transport towards lower energy sites within an energetically disordered landscape of the thin film.[36, 46, 47] The transient redshift of energy ($\Delta E$) for the three hybrid perovskite NC solids (C8, C12, and C16) is depicted in **Figure 3b** and follows an exponential decay of the form of $\Delta E(t) = \Delta E_{\infty}[1 - \exp(-k_{\Delta E}t)]$, where $\Delta E_{\infty}$ [meV] is the saturation energy and $k_{\Delta E}$ [$s^{-1}$] represents the rate of the energetic redshift.[36] The fitting parameters are displayed in **Figures 3c** and **3d**, showing that both the rate of redshift, $k_{\Delta E}$, and its amplitude, $\Delta E_{\infty}$, increase with decreasing $d_{C-C}$. The observation of increased energetic disorder despite the more monodisperse ensemble of the smaller NCs (see **Figure 1c**) may seem counterintuitive. To understand this apparent disconnect between structural and energetic disorder, it is important to consider the degree of quantum confinement of the excitonic excited state. While the larger particles (shorter alkyl chains) exhibit bandgap energies closer to those of bulk $CH_3NH_3PbBr_3$,[48, 49] the smaller particles (longer alkyl chains) show a clear blueshift in emission, where quantum confinement amplifies the variation in the effective bandgap, even for small variations in the particle size.[21]

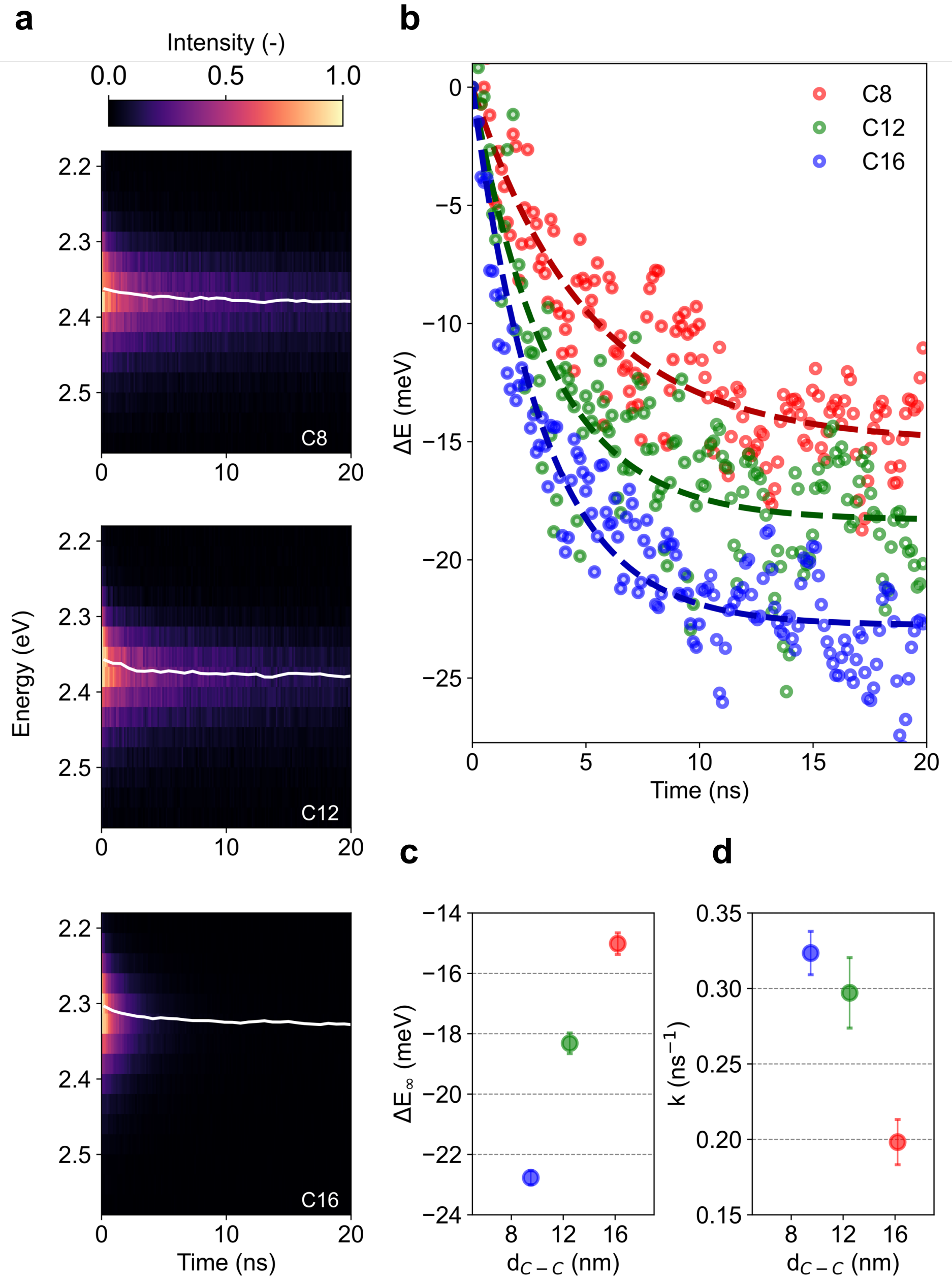

Figure 3. Energetic disorder in the hybrid LHP NC solids. (a) Spectrally resolved transient PL maps of $CH_3NH_3PbBr_3$ NC solids synthesized with C8, C12, and C16, respectively. The linear color scale represents the normalized PL intensity. The solid white line is the smoothed median emission wavelength. (b) Redshift of median emission energy, $\Delta E$, as a function of time. Dashed lines are fits to $\Delta E(t) = \Delta E_{\infty}[1 - \exp(-k_{\Delta E}t)]$. Fits are performed to the first 10 ns. (c) Saturation energy, $\Delta E_{\infty}$, and (d) rate of the energetic redshift, $k_{\Delta E}$, extracted from the fits in panel b, as a function of the center-to-center distance ($d_{C-C}$).

To understand the influence of the respective structural and energetic disorder in the different hybrid perovskite NC solids on exciton transport, we perform transient photoluminescence microscopy (TPLM), allowing for a direct visualization of exciton transport with few-nanometer and sub-nanosecond resolution. Diffusion of excitons is measured using a home-built microscope setup.[50] Briefly, we use a high-magnification (100× 1.45 NA) objective to focus the excitation laser ($\lambda_{exc}$ = 405 nm) to a near-diffraction-limited spot. The PL emission is then projected using a relay system (total magnification 360×) outside the microscope, where we raster scan the image using a motorized stage with a detector (APD, avalanche photodiode). **Figure 4a** demonstrates the resulting TPLM map with the evolution of the PL emission intensity in space and time, $I(x,t)$, for the hybrid NC solid synthesized with C8 (see **Figure S7** for diffusion maps for C12 and C16). To highlight the broadening of the exciton distribution within the film, $I(x,t)$ is normalized at each point in time to the maximum of the spatial profile.[51] We quantify the temporal broadening of the exciton distribution by fitting each time slice to a Voigt function to obtain a value of the variance ($\sigma^2$) as a function of time (for details, see Supporting Information).[50] We plot the change in $\sigma^2$ or the mean square displacement, MSD $= \sigma^2(t) - \sigma^2(t=0)$, in **Figure 4b**. We observe a clear correlation between the exciton distribution extent and the alkyl chain length of the ligand, with more pronounced diffusion for shorter alkyl chain lengths (larger NCs).

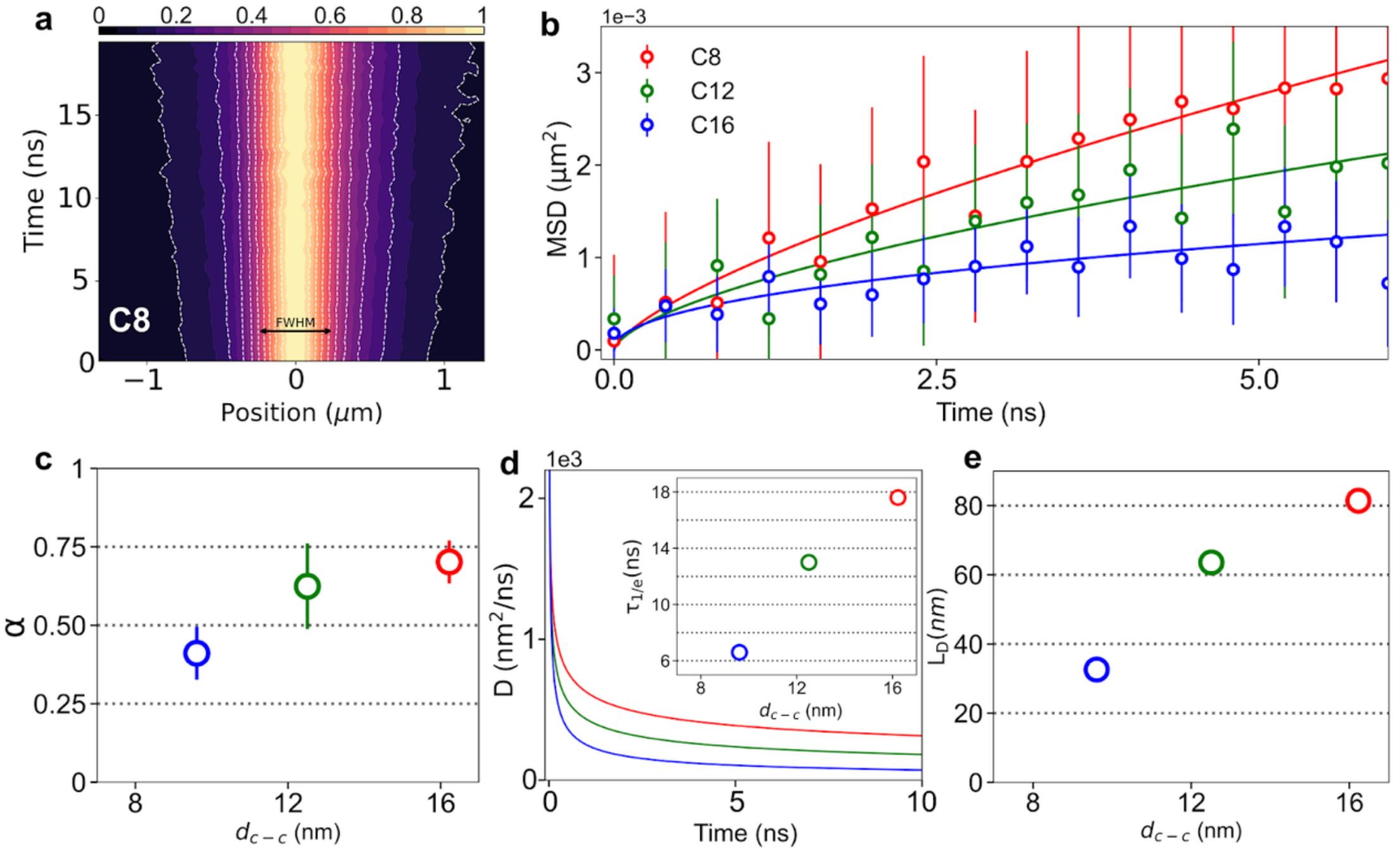

Figure 4. Transient PL microscopy of hybrid LHP NC solids. (a) Normalized TPLM map showing the spatial and temporal evolution of the PL emission intensity, $I(x,t)$, for the C8 system. (b) Evolution of the exciton distribution width as a function of time for all measured compounds (C8, C12, and C16). The width is calculated using the mean square displacement, MSD, of the population at each time position. (c) Extracted diffusion exponent, α, as a function of $d_{C\text{-}C}$. (d) Time-dependent diffusivity, $D$, for each compound. Inset shows the extracted exciton lifetime, $\tau_{1/e}$, obtained from the TPLM decay curves. (e) Calculated diffusion length, $L_D$, for all measured hybrid perovskite NC films.

Importantly, the diffusion dynamics of all three systems follows a clear non-linear evolution. We can quantify the different exciton dynamics in these systems by using the one-dimensional diffusion equation as a simple approximation. It follows that $\mathrm{MSD} = 2Dt^{\alpha}$, where $D$ [nm$^2$ ns$^{-1}$] is the diffusivity and α the diffusion exponent.[50] These fits are shown in Figure 4b as solid lines, with α is depicted as a function of $d_{C\text{-}C}$, in **Figure 4c**. If α = 1, normal diffusion takes place through conventional random-walk behavior and the exciton hopping rate is time-independent. For α < 1, subdiffusive transport of the charge carriers with a time-dependent diffusivity is observed.[52] For all samples, we find subdiffusive transport with a progressively decreasing α with decreasing $d_{C\text{-}C}$. The resulting time-dependent diffusivity (**Figure 4d**) is extracted by taking the derivative of the obtained fits in Figure 4b. Subdiffusive transport is indicative of an energetically disordered landscape, where asymmetric site-to-site hopping rates result in reduced exciton mobilities as carriers reach the valleys of the energy landscape.[37, 53, 54] The smaller α for smaller particles (C16) is therefore consistent with increasing energetic disorder that we also observe in the spectrally resolved transient PL measurements (see Figure 3**)**.

We note that subdiffusive behavior may also originate from trap-state limited diffusion.[50, 52] To verify the origin of the variation in α, we look at the PLQY and calculate the radiative and non-radiative rates by taking the respective lifetimes into account (see Table S1). We consistently observe a reduced PLQY when going from solution to thin film, which is accompanied by an increase in non-radiative rate ($k_{nr}$) across all three ligand systems, despite their significantly different exciton diffusivities (C8 > C12 > C16). In thin films, the extracted $k_{nr}$ is large and comparable across C8–C16, leading to similar lifetimes (6–8 ns) and PLQY values (~50%), which suggests that the quenching mechanism is local (short-range) rather than diffusion-limited. If exciton migration to trap sites were responsible for the PLQY loss, C8 would be expected to show a substantially larger $k_{nr}$ increase than C16. Notably, the $k_{nr}$ values in solution are low across C8 and C12, and even reduced for C16, confirming that all three ligand systems provide strong surface passivation, and that differences in the solution PLQY can be attributed to changes in the radiative rate ($k_r$) due to differences in the degree of quantum confinement. Based on this, we can conclude that the differences in α originate from a site-to-site disordered energy landscape.

To determine the exciton diffusion length, $L_D$, of the different systems (**Figure 4e**), we take into account the PL lifetime decay and the time-dependent diffusivity (see **Figure S8** for more details). The combination of a longer lifetime (see inset of Figure 4d) and increased diffusivity leads to a significantly longer $L_D$ for the longer $d_{C\text{-}C}$ (shorter alkyl chain system, C8), increasing from 30 nm for C16 to 64 nm for C12 and further up to 81 nm for C8. These values are comparable or even higher compared to conventional disordered metal chalcogenide semiconductor NC solids.[37, 55] At the same time, it is important to mention that our calculated values for $L_D$ do not reach those reported for superlattice and closed-packed assembly systems (*e.g.* $L_D$ = 200 nm for $CsPbBr_3$ NC assemblies) in which structural and energetic disorder are largely absent.[29, 56]

Our observation that larger and less ordered particles lead to faster exciton transport is surprising at first. According to Förster theory, smaller NCs are expected to exhibit stronger dipole–dipole coupling and, consequently, more efficient exciton transport. First, dipole-dipole coupling depends on the center-to-center distance between neighboring NCs rather than the interparticle spacing, which inherently favors smaller NCs. Second, the coupling strength scales with the exciton oscillator strength, which is increased under stronger quantum confinement and therefore also benefits smaller NCs. Consistent with this expectation, we indeed observe faster radiative rates for the smaller NCs (see Table S1). Taken together, all relevant considerations within the Förster framework predict that exciton transfer should improve with smaller NC sizes that are obtained with longer alkyl chain lengths. Strikingly, our direct visualization of exciton transport reveals the opposite trend: increasing alkyl chain length leads to progressively slower transport exciton dynamics. This apparent contradiction demonstrates that exciton transport is not limited by the film morphology, but instead by energetic disorder. In smaller, more strongly confined nanocrystals, size polydispersity leads to larger bandgap variations and greater site-to-site energetic disorder, which in turn enhances subdiffusive transport and suppresses exciton diffusion.

**Figure 5** confirms this behavior using numerical simulations of exciton transport (for technical details, see Figure S9 and Table S2). Taking the NC size distributions (Figure 1d) and the absorption maxima (Figure 2c) of the different samples, we obtain an empirical relationship between NC size and the effective optical bandgap (**Figure 5a**), allowing us to construct model landscapes for the different thin films with corresponding energetic disorder (**Figure 5b**). Assuming a $d^{-6}$ distance scaling for energy transfer and $E^{-\Delta E/k_B T}$ for the energetic difference, we can reproduce both the relevant trends in the transient redshift (**Figure 5c**) and the subdiffusive spatial dynamics (**Figure 5d**) for the different thin films.

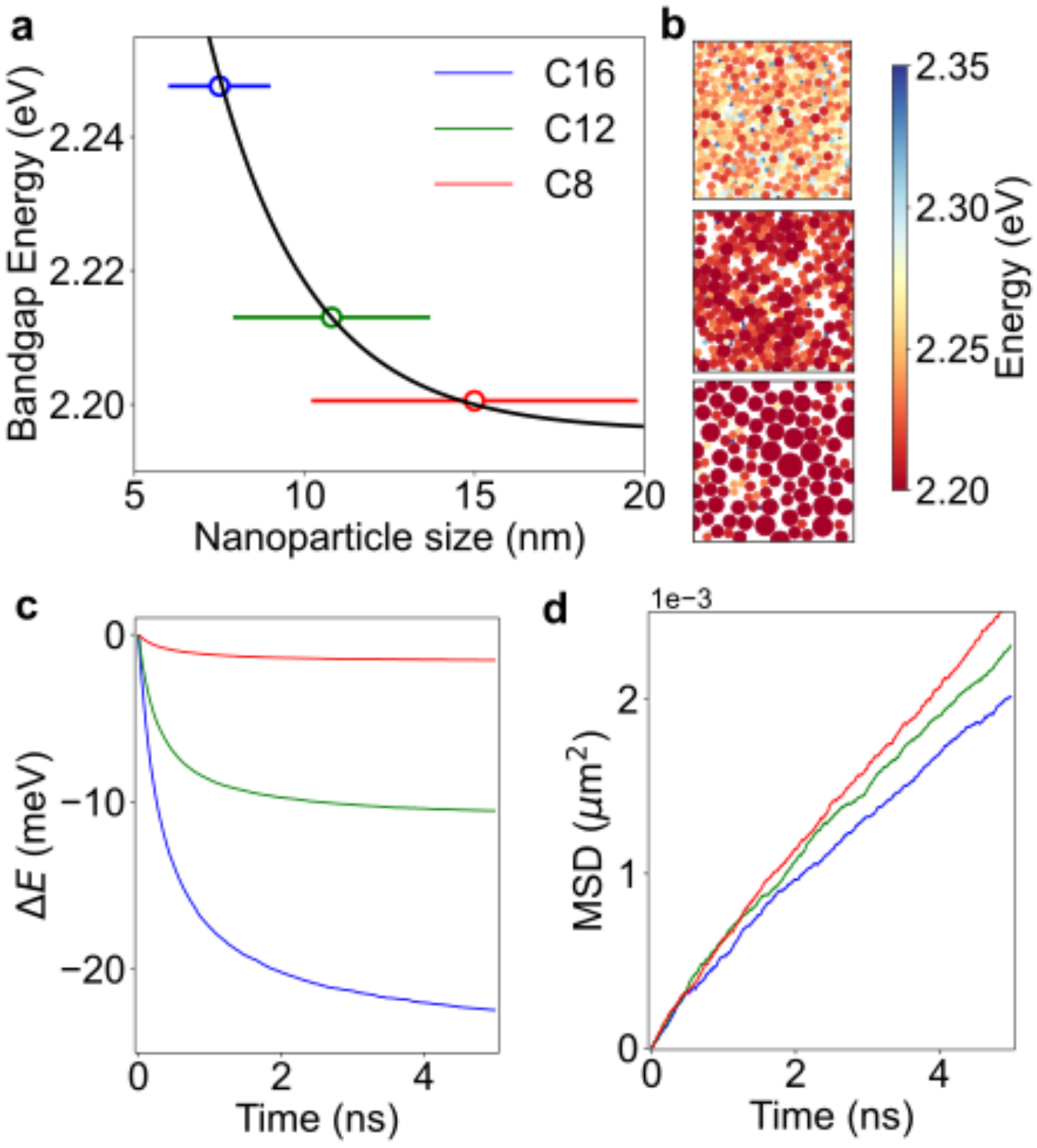

Figure 5. Numerical simulations of exciton transport in disordered NC thin films. (a) Bandgap energy (see Figure 2) as a function of average NC size (see Figure 1). Horizontal error bars represent the experimentally obtained polydispersity of the NCs. Solid black line is an empirical function, given by $E(d) = 0.70254 \cdot \exp(-0.3446 \cdot d) + 2.196$. (b) Random two-dimensional model lattices with structural and energetic disorder generated based on (a). (c) Simulated redshift of median emission energy, $\Delta E$, as a function of time. (d) Evolution of the exciton distribution width as a function of time for all simulated lattices (C8, C12, and C16).

In conclusion, our study shows the complex relationship between structural and energetic disorder in LHP NC solids. In the specific case of room-temperature synthesized organic-inorganic hybrid $CH_3NH_3PbBr_3$ NCs, we observe that the choice of the amphiphilic alkylamine ligand used for the colloidal synthesis of the LHP NCs greatly affects the extent of disorder. When it comes to exciton diffusion, the amount of energetic disorder dominates over the presence of structural disorder, resulting in more subdiffusive behavior and less efficient energy transport upon decreasing NC sizes for longer alkyl chain lengths. Our results emphasize the importance of the type of disorder present in semiconductor NC solids and the right selection of ligands to rationally design NCs solids with the desired energy transport properties.

**ACKNOWLEDGEMENTS**

This work was funded by the European Union (ERC, EnVision, project number 101125962). Views and opinions expressed are however those of the author(s) only and do not necessarily reflect those of the European Union or the European Research Council Executive Agency. Neither the European Union nor the granting authority can be held responsible for them. F.P. further acknowledges funding from the Spanish AEI under grant agreements PID2022-141579OB-I00, TED2021-131018B-C21, and CNS2023-143577 and the "María de Maeztu" Program for Units of Excellence in R&D (CEX2023-001316-M). In addition, we acknowledge the support from the "(MAD2D-CM)-UAM" project funded by Comunidad de Madrid, by the Recovery, Transformation and Resilience Plan, and by NextGenerationEU from the European Union. A.T.-P. thanks the Comunidad de Madrid for funding STEM experiments under the PR65/19-22438 Research Project. B. H. J. acknowledges the project PID2023-151371OB-C21 by MCIN/AEI/10.13039/501100011033 and by FEDER, EU and the Severo Ochoa Centres of Excellence program through Grant CEX2024-001445-S.

**AUTHOR CONTRIBUTIONS**

S.F.S. and F.P. conceived the idea and designed the experiments. S.F.S. performed the synthesis, fabricated the thin films, and measured absorption. E.A.R. measured PL and performed and analyzed TPLM experiments. A.Cu. measured and analyzed spectrally resolved lifetimes. A.Co. and B.H.J. assisted in the synthesis and structural analysis, measured and analyzed PL lifetimes, and performed quantum yield measurements with assistance of A.D.A.. M.M. conducted numerical simulations. A.T.-P. acquired STEM images and performed size analysis of the NCs together with A.Co. and B.H.J.. S.F.S., B.H.J., and F.P. co-wrote the manuscript. All authors contributed to this work, read the manuscript, discussed the results, and agreed to the contents of the manuscript and supporting information.

**DATA AVAILABILITY**

The data supporting the findings of this study are available within the article and its Supplementary information.

**CODE AVAILABILITY**

Correspondence and requests for codes used in the paper should be addressed to the corresponding author.

**ETHICS DECLARATIONS**

The authors declare no competing interests.

# Supporting Information

## Exciton Transport in Disordered Perovskite Nanocrystal Solids

Simon F. Solari[1,2], Enrique Arévalo Rodríguez[1,2], Antonella Cutrupi[1,2], Amalia Coro[1,2,3], Marc Meléndez[1,2], Alicia De Andrés[3], Almudena Torres-Pardo[4], Beatriz H. Juárez[3], and Ferry Prins[1,2*]

[1] Condensed Matter Physics Center, IFIMAC, Universidad Autónoma de Madrid, Madrid 28049, Spain

[2] Departamento de Física de la Materia Condensada, Facultad de Ciencias, Universidad Autónoma de Madrid, Madrid 28049, Spain

[3] Instituto de Ciencia de Materiales de Madrid, ICMM, Consejo Superior de Investigaciones Científicas, CSIC, Madrid 28049, Spain

[4] Departamento de Química Inorgánica, Facultad de Ciencias Químicas, Universidad Complutense de Madrid, Madrid 28040, Spain

*All correspondence should be addressed to ferry.prins@uam.es

## Experimental Section

### Synthetic protocol

***Chemicals:*** Lead(II) bromide ($PbBr_2$, ≥98%, Merck), Methylammonium bromide ($CH_3NH_3Br$, 98%, Merck), *N*,*N*-Dimethylformamide (DMF, ACS reagent, ≥99.5%, Merck), Ethanol (EtOH, absolute for analysis, Merck), Toluene (ACS reagent, ≥99.5%, Merck) Oleic acid (OLAc, technical grade, 90%, Merck), Octylamine (C8, 99%, Merck), Dodecylamine (C12, ≥99%, Merck), Hexadecylamine (C16, 98%, Merck), Methyl acetate (MeOAc, Scharlab). All chemicals listed above were used as received without any further purification.

***Synthesis of $CH_3NH_3PbBr_3$ nanocrystals (NCs):*** The synthetic protocol for the fabrication of the colloidal $CH_3NH_3PbBr_3$ NCs was slightly modified from a previous report by Kumar *et al.* from 2019.[1] Briefly, $PbBr_2$ was dissolved in DMF (0.6 M), while $CH_3NH_3Br$ was dissolved in EtOH (0.6 M). OLAc (625 µL) and the amphiphilic ligand (25 µL C8, 35 µL C12, or 36.4 mg C16) were mixed with toluene (12.6 mL) in a 100 mL one-neck flask. Afterwards, the reaction contents were connected to a Schlenk line to purge the system three times with argon. Although the colloidal synthesis of the NCs could be performed under ambient conditions, it has been observed that the quality and the yield of the colloidal perovskite NCs varied with respect to the outer conditions such as air humidity.[2] After purging, the $PbBr_2$ (625 µL, 0.6 M, in DMF) and the $CH_3NH_3Br$ (625 µL, 0.6 M, in EtOH) precursor solutions were added to the reaction mixture, followed by another three purging cycles with argon. Subsequently, the reaction contents were separated by means of centrifugation (8000 rpm, 7227 *g*, 8 min) and the precipitate was redispersed in 3 mL of fresh toluene. Afterwards, the reaction contents were centrifuged again (8000 rpm, 7227 *g*, 8 min). The supernatant was then filtered through a PTFE syringe filter (pore size = 0.22 µm) and directly transferred to a new centrifuge tube. 2 mL of MeOAc were added to this centrifuge tube for the purification of the $CH_3NH_3PbBr_3$ NCs. The reaction mixture was separated employing centrifugation (8000 rpm, 7227 *g*, 8 min) and the precipitate was redispersed in 1 – 1.5 mL of fresh toluene, followed by another centrifugation cycle (8000 rpm, 7227 *g*, 8 min). Finally, the resulting solution was filtered through a PTFE syringe filter (pore size = 0.22 µm). The green solution containing colloidally dispersed $CH_3NH_3PbBr_3$ NCs was stored for further experiments.

**Fabrication of perovskite thin films**

***Procedure:*** For the fabrication of the perovskite NC solids, thin glass substrates (24 mm × 24 mm, 0.13 – 0.16 mm thickness) were first sonicated in isopropanol for a few minutes and then cleaned for 5 min with a UV Ozone Cleaner (Ossila). Afterwards, the cleaned substrates were transferred to a glovebox filled with $N_2$ and mounted on a substrate holder of a Spin Coater (Ossila). 100 µL of the synthesized colloidal NC dispersion were dropped onto the substrate, followed by spin-coating at 2500 rpm for 40 seconds. The resulting perovskite thin films were stored for further experiments. We note that some variation in the film thickness between different samples is present due to variations in the nanoparticle concentration after synthesis. This, however, has no significant influence on the results.

## Characterization

***Absorption spectroscopy:*** UV-visible absorption spectra of the liquid NC dispersions were measured using a Varian Cary 50 Conc UV-VIS Spectrophotometer. The liquid samples were approximately diluted up to 200 times with toluene. The absorption band edge was determined by calculating the first derivative of the corresponding interpolated and smoothed absorption curves.

***Photoluminescence Quantum Yield (PLQY) measurements:*** Emission lifetimes and corresponding absolute photoluminescence quantum yields (PLQY) of colloidal solutions and thin films were measured with an FLS1000 fluorescence spectrometer (FLS1000 Fluorometer, Edinburgh Instruments). The instrument is equipped with high-throughput double monochromators in both excitation and emission paths, single-photon counting detectors, and time-correlated single-photon counting (TCSPC) electronics for lifetime measurements. Absolute PLQY measurements were carried out using a calibrated integrating sphere accessory integrated into the system, allowing direct determination of quantum yield without the need for external standards.

***Photoluminescence Spectroscopy:*** PL spectra were obtained in a home-made system where samples were mounted on an inverted microscope (Nikon Eclipse Ti-U). To obtain PL spectra, the perovskite films were excited using either a 405 nm laser (PicoQuant LDH-D-C-405, PDL 800-D) or a 385 nm light-emitting diode (Thorlabs, M385LP1-C5). Emission spectra were recorded using an imaging spectrograph (Princeton Instruments, SpectraPro HRS-300, ProEM HS 1024BX3). The setup was additionally connected to a Thorlabs scientific CMOS camera (CS165CU/M) to acquire film images.

***Power-dependent PL spectroscopy:*** The excitonic nature of the carriers was determined by PL power-dependent measurements. The samples were excited using a laser diode (PicoQuant LDH-D-C-405, PDL 800-D). A low magnification objective (Nikon Plan Fluor 10x) is selected along with a defocused laser spot to minimize laser fluence and reduce potential Auger effects happening at high illumination intensities. The excitation power is controlled by using a continuous ND filter wheel (Thorlabs NDC-100C-4M-A) to not change the shape of the excitation spot when changing powers. The emitted PL signal is sent to a photodetector (avalanche photodiode (APD), Micro Photon Devices PDM) which collects light for different values of the laser intensity. The excitation intensity is measured before the objective using a photodiode power sensor (S120C). The counts obtained in the APD are summed together to obtain a total intensity value for each laser power.

***Transient photoluminescence microscopy (TPLM):*** Exciton diffusion was measured following the same procedure as Akselrod *et al.* from 2014.[3] Briefly, a near diffraction limited laser (PicoQuant LDH-D-C-405, PDL 800-D) is used to excite the sample. The emission light is then collected using a x100 oil immersion objective (Nikon Plan APO λD, NA = 1.45). The PL image is projected using a large magnification (360x) relay system onto the photodetector (APD, Micro Photon Devices PDM). Laser and APD are synchronized by means of a time correlated single photon counting device (PicoHarp 300). The APD was scanned through the center of the excitation spot by using two linear actuators (Thorlabs Z912). All TPLM measurements were performed using a laser fluence of 21 nJ/cm$^2$ and 20 MHz repetition rate. To prevent photo-induced degradation and account for spatial inhomogeneities within the films, TPLM measurements were performed at multiple positions for each compound. The collected data was merged to improve signal-to-noise ratio. Accordingly, the spot-to-spot variation in diffusivity is reflected in the error bars presented in Figure 4. Data acquisition and analysis were implemented and done using custom python code. The TPLM data is fitted to a Voigt function to extract the value of the variance of the Gaussian part of the Voigt ($\sigma^2$) and the full width at half maximum (FWHM) of the Lorentzian. Fitting the experimental data to a Voigt instead of a Gaussian better reproduces the broad tails of the profiles and can be understood as the convolution of the initial excitation spot, which was found to be a Lorentzian,[3] and a Gaussian that broadens over time. Consequently, the mean square displacement (MSD) can be calculated as $\text{MSD} = \sigma^2(t) - \sigma^2(t=0) = 2Dt^{\alpha}$. For material with high concentration of trap states, transport is subdiffusive ($0 < \alpha < 1$). For each sample, we find the value of α that best fits MSD(*t*). Time-dependent diffusivity, *D*(*t*), is then calculated by extracting the derivative of the MSD curve for all time points. To determine the exciton diffusion lengths, lifetime traces of all film samples were extracted from the TPLM results. Specifically, the decays were extracted by spatially integrating the counts along the diffusion scan range.

***Spectrally resolved lifetime measurement:*** Spectrally resolved transient photoluminescence was performed using a custom-built fluorescence microscope. The excitation source was a 375 nm pulsed laser diode (PicoQuant LDH-D-C-375, PDL 800-D; 5 MHz, 5.8 nJ cm$^{-2}$). The beam was coupled into a single mode fiber with a length of 1 m. The output beam was directed to an inverted optical microscope (Nikon Eclipse Ti-U) and the excitation laser was filtered out using a dichroic beamsplitter (SemRock FF376-Di01-25X36). A 100x oil immersion objective (Nikon CFI Plan Fluor, NA = 0.5 - 1.3) was used to focus the laser beam to a spot with a full width at half maximum (FWHM) of approximately 767 nm. The fluorescence beam was focused into the spectrograph (Princeton Instruments SpectraPro HRS-300 imaging spectrograph), diffracted by a grating of density 300 g/mm, blaze of 500 nm, and spectrally filtered by the exit slit (100 nm). The spectral line was focused on a single photon detecting avalanche photodiode (APD, Micro Photon Devices PDM, 20 × 20 µm detector). The laser and APD were synchronized using an electronic time-correlated single photon counting system (TCSPC), PicoHarp 300. The experimental data were collected using a custom Python script, which recorded a TCSPC histogram at each position as the spectrograph grating was scanned across the spectral range of 484 nm and 564 nm in 5.3 nm increments. The result was a map of spectrally resolved fluorescence intensity as a function of time.

***Scanning transmission electron microscopy (STEM):*** STEM experiments were performed at the Centro Nacional de Microscopía Electrónica (Madrid, Spain). For the STEM investigation, a droplet of the nanoparticle suspension was deposited on a thin carbon film supported on a Cu TEM grid. STEM investigations were performed on an imaging aberration-corrected microscope JEOL JEM GRAND ARM300cF operating at 300 kV. Images were recorded with a high-angle annular dark field (HAADF) detector. The beam intensity and the image acquisition time were optimized for each sample to prevent sample damage caused by the electron beam. For the histogram shown in Figure 1d, the main sizes of platelets were obtained as the arithmetic mean of the two side lengths. The statistics were done with 50 NCs in each case (C8, C12, and C16, respectively).

**Monte Carlo simulations of exciton diffusion in disordered solids**

To better understand how disorder affects exciton transport in perovskite nanocrystals we performed numerical simulations by modeling excitons as random walkers. In the case of mere structural disorder, the probability of a random walker moving to a nearby spot depends both on its current location and the direction, but the probability of moving from point A to point B equals that of moving back from B to A. The systems studied in our work also contain energetic disorder. In the latter case, walkers have a higher probability of moving towards lower energy locations than in the opposite direction.

We will assume that, though the probabilities change from point to point, the overall randomness of the landscape is homogeneous and isotropic. The concentration at any given point in the landscape remains constant whenever the number of walkers flowing into the area equals the number flowing out. On the one hand, in the case of structural diffusion, this occurs when a given area and its environment have the same concentration of walkers. Even though some points may be more difficult to reach initially (because the paths leading to them have lower probabilities), there are no favored points in the long run. Consequently, the distribution of transition probabilities that the walker encounters does not change with time.

On the other hand, if the landscape includes energetic disorder, walkers move downhill in energy on average. This means that, as time goes by, they encounter more and more transitions with lower probabilities, which in turn slows down their diffusion.

The former structural disorder landscape will lead to Brownian diffusion. To see why, let $\Delta r$ stand for a random displacement and p for a transition probability associated with such a displacement. Because p depends on the specific location of the walker, we represent the probability of encountering a given value for p as $P(p, \Delta r)$. Then, the probability of going from $r$ to $r + \Delta r$ equals

$$P(\Delta r) = \int P(p, \Delta r) p dp$$

The crucial point here is that, even though the distribution of values of $p$ may depend on the magnitude of $\Delta r$, it does not depend on time. The probability distribution $P(\Delta r)$ might differ dramatically from a Gaussian distribution.

However, as we accumulate displacements, each given by an identical independent distribution, the central limit theorem guarantees that the probability distribution for a large number of displacements, say *N*, will approach a Gaussian. Over longer time scales, such as *2N* displacements, the distribution for the total displacement will equal the convolution of two Gaussian functions, which is also a Gaussian with a variance equal to the sum of the two variances. This would make the variance proportional to the time, the hallmark of Brownian motion.

The same argument cannot be applied with energetic disorder because the distribution of $p$ values does depend on time in that case, as the populations of different energy levels change with time.

In summary, we expect Brownian diffusion in landscapes with structural disorder, but a gradual decrease in diffusivity when moving through materials with energetic disorder.

We performed numerical Monte Carlo simulations to verify that energetically disordered systems like those presented in our work lead to the observed subdiffusion. To model the systems, we start by preparing a polydisperse distribution of disks on a plane (see Fig. 5b in the main text), where the concentration of disks and their sizes were selected based on the TEM images (Fig. 1c,d). We introduced an initial population of $N = 5 \times 10^4$ excitons that are allowed to jump from one disk to nearby disks with a probability of:

$$p = \frac{1}{\left(\gamma(r+r_0)\right)^6} \quad (1)$$

following the expectation for dipoles. In equation 1, the $r_0$ term guarantees that the probability does not diverge as the distance $r$ tends towards zero, and $\gamma$ is a term determined by the transfer rate.

To simulate energetic disorder, we added an extra coefficient with the form $e^{-\Delta E/k_b T}$, where $\Delta E$ is the energy difference between donor and acceptor crystals, which makes it less likely for excitons to move uphill in energy. The energy of a site depends on the crystal size (diameter $d$) according to the following empirical function

$E(d) = Aexp(-bx) + E_b.$ (2)

The shape and parameter values of this expression were chosen to fit the measured bandgap energies of the different crystals (Fig. 5a). See Table S2 below for a full list of the parameter values.

We start the simulation with an initial Gaussian population of 200 nm standard deviation and centered at (0,0) for all three crystal structures shown in Fig 5b. For each exciton, the algorithm picks a random neighbouring crystal and determines whether the exciton moves to the new site by calculating the probability of transferring to the neighbour. This probability is either $p$ from Eq. (1) above (when the donor crystal is smaller than the acceptor) or it equals p times $e^{-\Delta E/k_b T}$ (when moving uphill in the energy landscape to a smaller crystal). If $d_1$ and $d_2$ stand for the diameters of the donor and acceptor sites, respectively, then the energy difference is calculated as

$\Delta E = E(d_2) - E(d_1) + N(0,2\sigma),$ (3)

where $N(0,2\sigma)$ represents a random number from a Gaussian distribution with standard deviation $2\sigma$, which models the uncertainty in the bandgap energies.

The process is repeated over many time steps and the total variance of the population and mean energy of the exciton states are tracked in time to create Figures. 5c, d.

**Evaluation of dielectric effects**

Regarding the possible role of dielectric effects in the energy transfer efficiency, the change in ligand length necessarily modifies the local dielectric environment between neighboring nanocrystals. However, a quantitative estimate shows that this effect is too small to account for the observed trends. Treating energy transfer within the Förster framework and approximating the donor-acceptor coupling through a distance-dependent effective medium, the effective refractive index experienced by the dipole-dipole interaction decreases only weakly when increasing the interparticle spacing from 1 to 2 nm (from an effective refractive index of approximately 1.96 to approximately 1.93 for a nanocrystal refractive index of 2 and ligand refractive index of 1.5). Since the Förster rate scales as the inverse fourth power of the refractive index ($n^{-4}$), this corresponds to a change in transfer rate of only a few percent, which is negligible compared to the change imposed by the inverse sixth power distance dependence ($R^{-6}$) and far smaller than the experimentally observed variations in diffusivity.

## Additional Figures

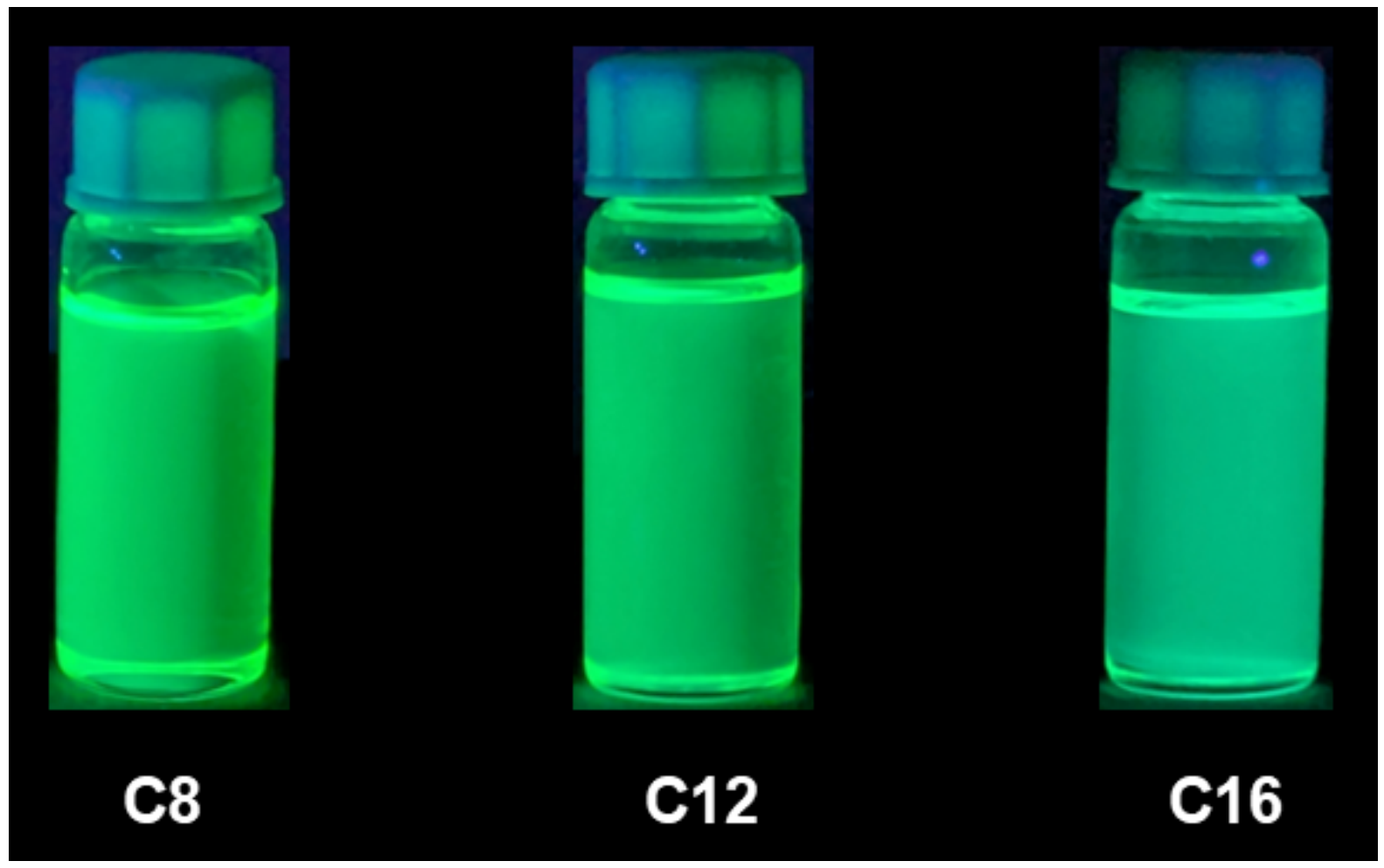

**Figure S1.** Images of the colloidal solutions of $CH_3NH_3PbBr_3$ NCs synthesized with C8, C12, and C16, respectively, under UV excitation (395 nm). The colloidal solutions exhibit strong green fluorescence.

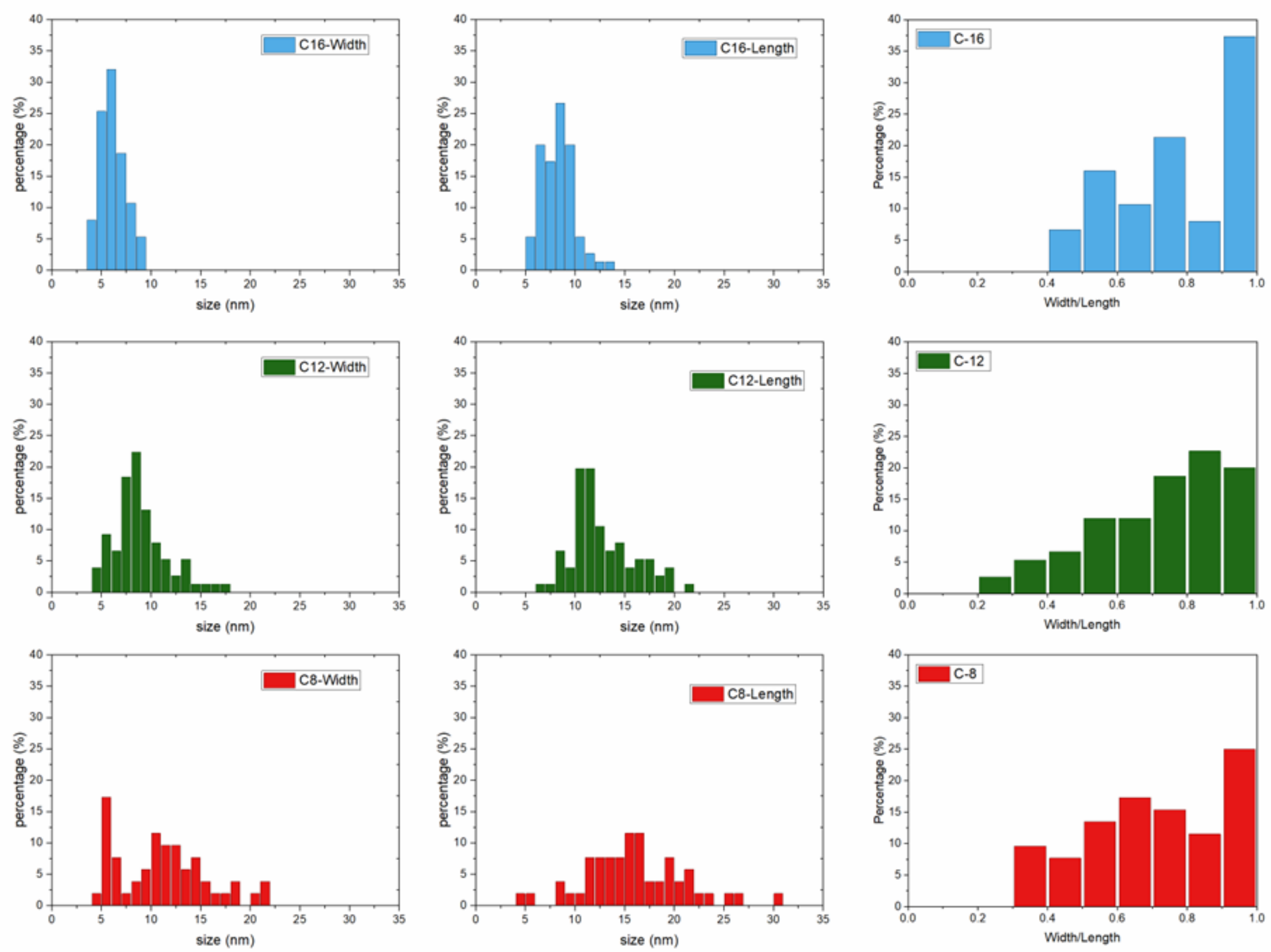

**Figure S2.** Statistical analysis of width (nm), length (nm), and aspect ratio for samples C16 (blue), C12 (green), and C8 (red). The analysis was carried out on 75–100 particles. The smallest size was identified for C16, which also shows a high percentage of particles (∼35%) with an aspect ratio close to unity (square-shaped). For samples C12 and C8, the aspect ratio deviates from unity, with a broader size distribution and an increasing average size.

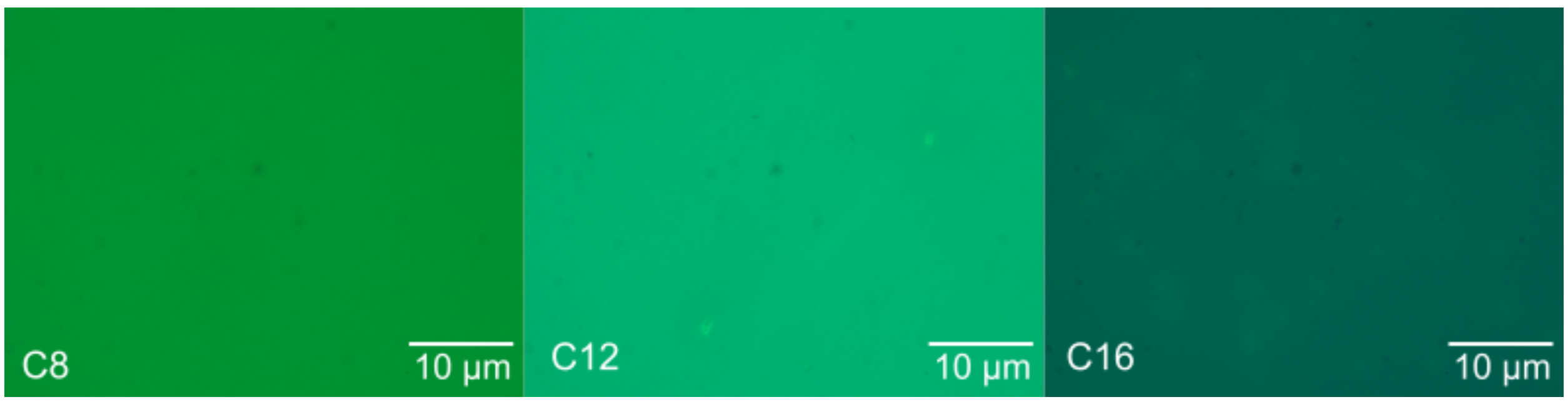

**Figure S3.** Images of the spin-coated perovskite thin films under LED illumination.

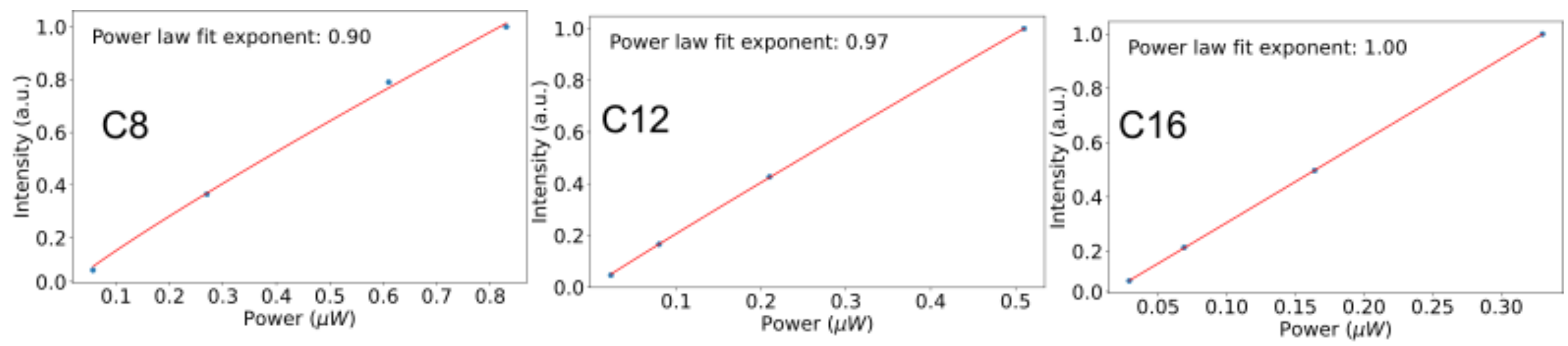

**Figure S4.** Results of the power-dependent PL measurements.

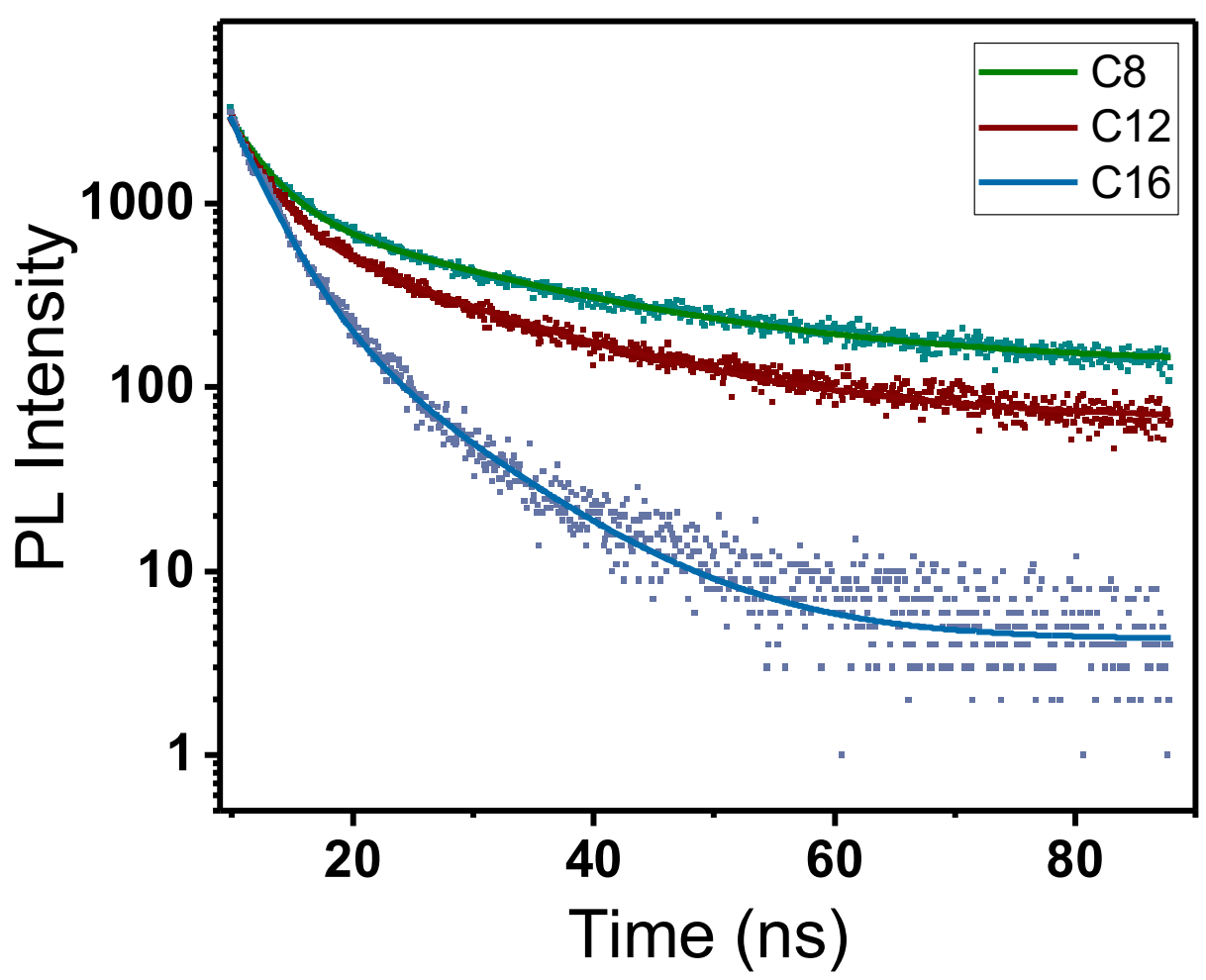

**Figure S5.** Lifetime measurements of of C8 (green), C12 (brown) and C16 (blue) colloidal NCs solutions. Systematic analysis of the curves revealed that the bi-exponential model provides a good description of decay kinetics ($R^2$ >0.99) described by $I(t) = A_1 exp(-t/\tau_1) + A_2 exp(-t/\tau_2)$. The intensity-weighted average lifetime was calculated according to $\tau_{avg} = \frac{A_1 \tau_1^2 + A_2 \tau_2^2}{A_1 \tau_1 + A_2 \tau_2}$ and included

in Table S1, along with all fitting parameters $A_i$ and $\tau_i$, and the radiative and non-radiative contributions according to the following expressions: $k_{nr} = \frac{1-PLQY}{\tau_{avg}}$, where $k_{nr}$ and $k_r$ refers to non-radiative rate and radiative rates, respectively $k_r = \frac{PLQY}{\tau_{avg}}$.

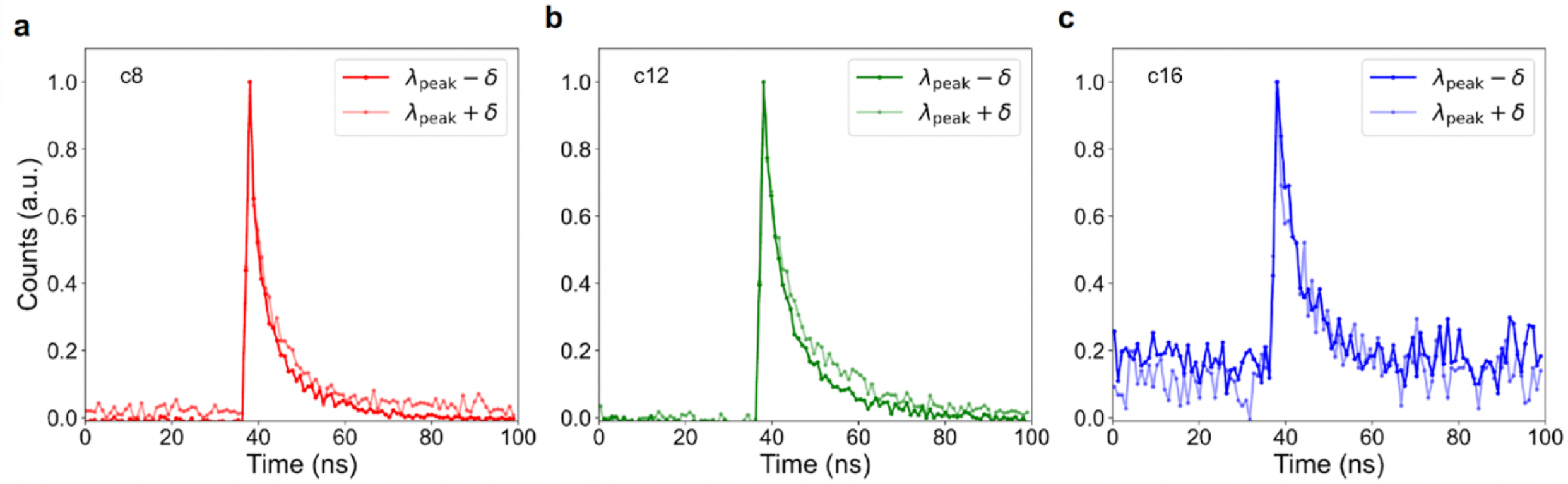

**Figure S6.** Spectrally resolved transient PL lifetimes traces of $CH_3NH_3PbBr_3$ NC liquid solutions synthesized with C8 (a) C12 (b) and C16 (c). The reported traces were collected at wavelength regions on the blue and red side of the maximum of the PL emission (δ = 20 nm).

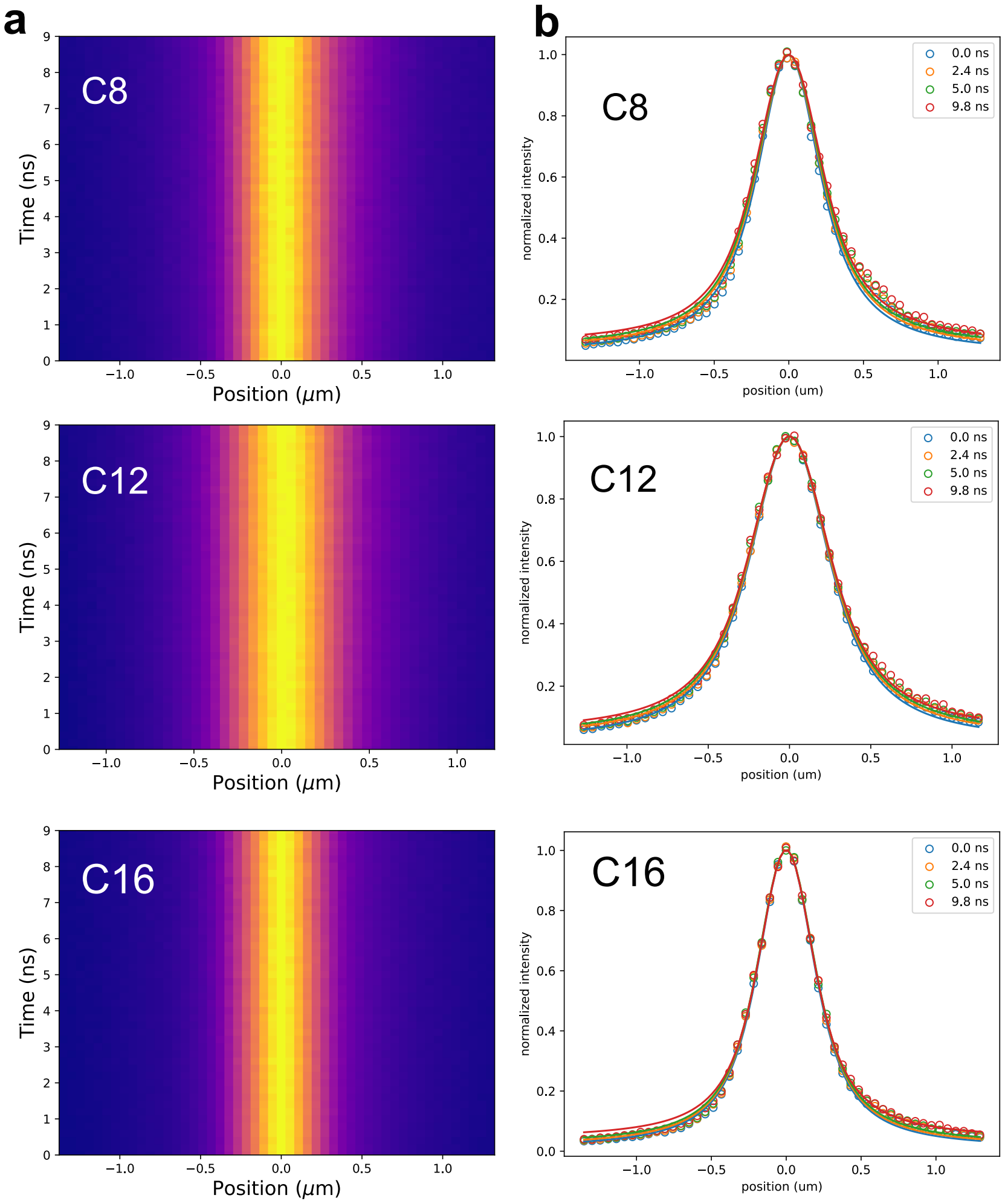

**Figure S7.** a) Diffusion maps of all samples measured. b) PL profiles and Voigt fits for selected times for each sample. Please note that while the width of the initial population at $t = 0$ can vary from measurement to measurement, this has no influence on the determination of the diffusivity since MSD $= \sigma^2(t) - \sigma^2(t = 0)$.

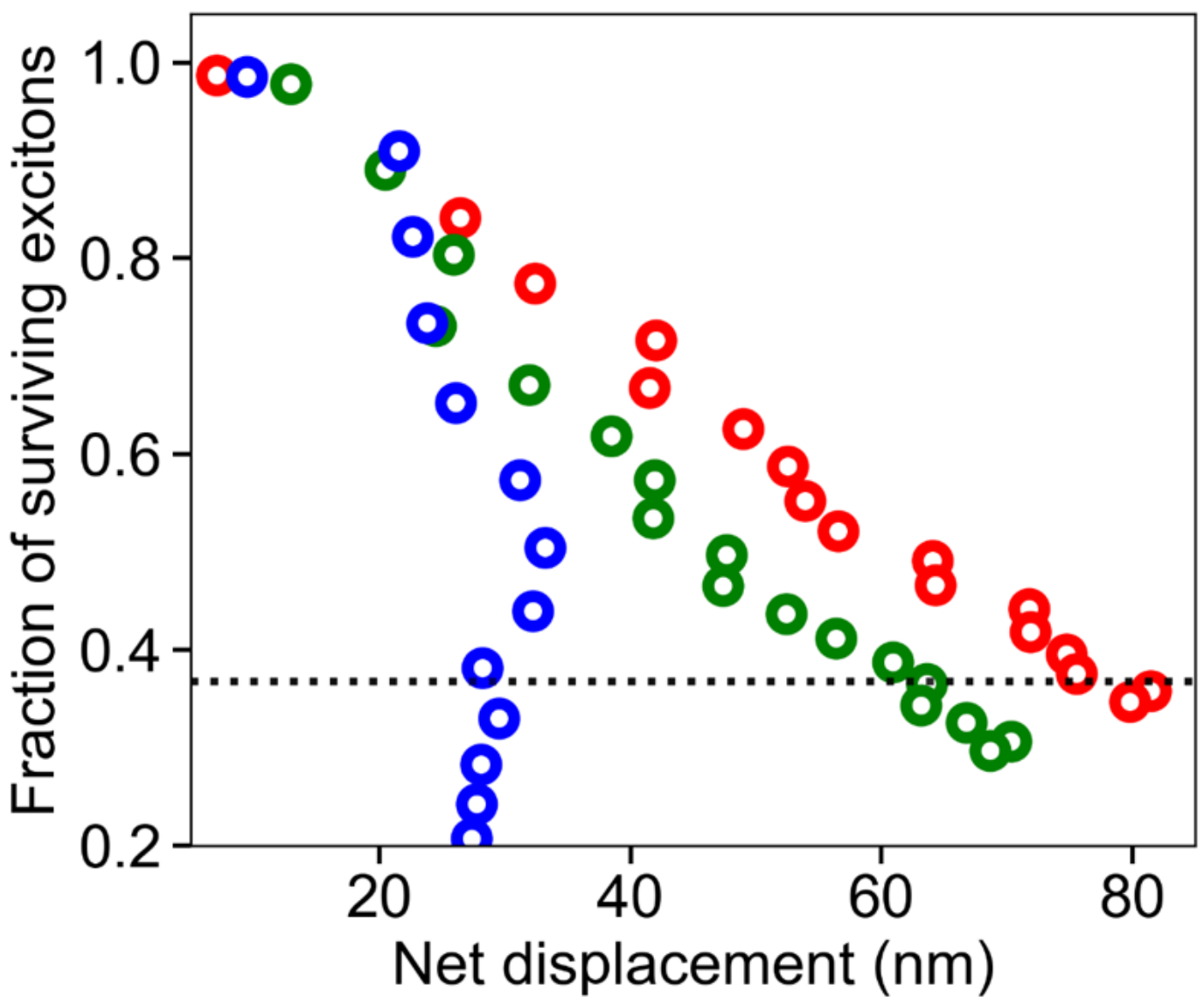

**Figure S8.** Fractions of surviving excitons as a function of net spatial displacement $\sqrt{\mathrm{MSD}(t)}$ of excitons for C8 (red), C12 (green), and C16 (blue), respectively. Diffusion lengths are calculated by estimating the fraction of surviving excitons from the integrated intensities. We use the net displacement of the population at each time position ($\sqrt{\mathrm{MSD}(t)}$) and find the value of the net displacement when the remaining population becomes 1/e of the initial one. The dashed line indicates the point when then population becomes 1/e of the initial one, the x value where each curve intersects with the dashed line corresponds to the diffusion length of each compound.[4]

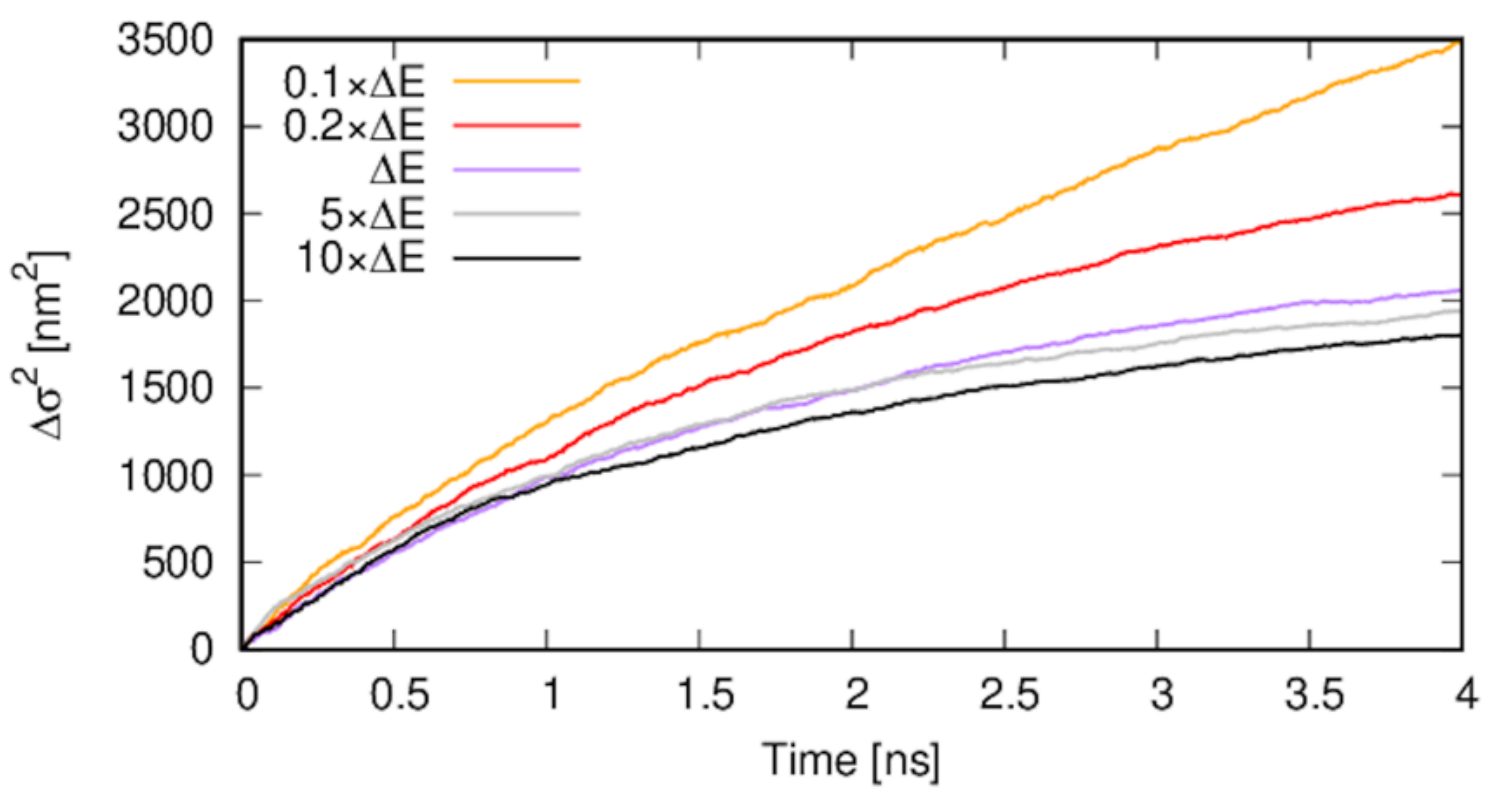

**Figure S9.** Simulated diffusion dynamics in a hypothetical system in which the centre-to-centre distances are kept constant but the degree of energetic disorder is varied by multiplying all the energy differences by a constant factor (see legend). Each line corresponds to a different degree of energetic disorder. As expected, α scales inversely with the degree of energetic disorder.

## Additional Tables

**Table S1.** QY values and charge carrier lifetimes of samples measured in solution and film. In solution, the PLQY increases systematically with ligand length, from 0.77 (C8) to 0.84 (C12) and 0.97 (C16). The time-resolved PL analyses yield mean PL lifetimes of 15.1 ns (C8), 11.4 ns (C12), and 4.9 ns (C16). Importantly, the combined PLQY–lifetime analysis for colloidal solutions allows us to separate radiative and nonradiative recombination contributions. As can be seen, the radiative recombination rate ($k_r$) increases by a factor of 4 from C8 (0.05 $ns^{-1}$) to C16 (0.2 $ns^{-1}$). In contrast, the nonradiative rate ($k_{nr}$) remains comparable for C8 and C12, and is slightly reduced for C16. The shorter lifetime from sample C8 to C16 (from approx. 15 to 5 ns) is driven predominantly by an increase in the radiative rate (faster radiative recombination, larger $k_r$ in C16 than in C8). This comparison allows confirmation that the observed PLQY enhancement with increasing ligand length is not driven by an increase in trap-assisted recombination; instead, it is dominated by a pronounced increase in the radiative rate (faster radiative recombination), consistent with enhanced radiative strength for the C16 sample under our synthesis/passivation conditions.

| Parameter | C8 (Sol) | C8 (Film) | C12 (Sol) | C12 (Film) | C16 (Sol) | C16 (Film) |
|---|---|---|---|---|---|---|
| QY | 0.77 | 0.51 | 0.84 | 0.48 | 0.97 | 0.52 |
| $A_1$ | 2.09 | 0.27 | 2.23 | 0.29 | 2.71 | 0.27 |
| $\tau_1$ (ns) | 2.94 | 1.54 | 2.70 | 1.38 | 2.82 | 2.03 |
| $A_2$ | 0.84 | 0.49 | 0.75 | 0.48 | 0.40 | 0.71 |
| $\tau_2$ (ns) | 19.62 | 12.00 | 15.78 | 10.45 | 9.20 | 8.24 |
| $\tau_{avg}$ (ns) | 15.10 | 8.28 | 11.40 | 7.03 | 4.90 | 6.53 |
| $k_r$ ($ns^{-1}$) | 0.051 | 0.062 | 0.074 | 0.068 | 0.198 | 0.080 |
| $k_{nr}$ ($ns^{-1}$) | 0.015 | 0.059 | 0.014 | 0.074 | 0.006 | 0.074 |

**Table S2.** Numerical Monte Carlo simulation parameter values.

| Parameter | Value | Meaning |
|---|---|---|
| $\gamma$ | 0.06 $nm^{-1}$ | Parameter in probability function Eq (1) determining transfer rate. |
| $r_0$ | 15 nm | Offset in Eq (1) that avoids singularity in the probability. |
| A | 702.54 meV | Bandgap energy fit parameter |
| b | 0.3446 $nm^{-1}$ | Bandgap energy fit parameter (decay with distance) |
| $E_b$ | 2196 meV | Bandgap energy fit parameter (bulk value) |
| $\sigma$ | 1.96 meV | Uncertainty in bandgap energy |

## Additional References